\documentclass[aps,prd,epsf,showpacs,amsmath,amssymb,graphics,preprint]{revtex4}
\usepackage{graphicx}
\usepackage{dcolumn}
\usepackage{bm}

\newcommand{\be}{\begin{equation}}
\newcommand{\ee}{\end{equation}}
\newcommand{\ba}{\begin{eqnarray}}
\newcommand{\ea}{\end{eqnarray}}
\newcommand{\ban}{\begin{eqnarray*}}
\newcommand{\ean}{\end{eqnarray*}}

\newcommand{\eq}[1]{(\ref{#1})}

\begin{document}
\title{Ultra-high energy collision with neither black hole nor naked singularity}
\author{$^1$Ken-ichi Nakao\footnote{Electronic address: knakao@sci.osaka-cu.ac.jp}
$^2$Masashi Kimura\footnote{Electronic address: mkimura@yukawa.kyoto-u.ac.jp}, 
$^3$Mandar Patil\footnote{Electronic address: mandarp@tifr.res.in},
$^3$Pankaj S. Joshi\footnote{Electronic address: psj@tifr.res.in}
}

\affiliation{$^1$Department of Mathematics and Physics, Graduate
School of Science, Osaka City University, Osaka 558-8585, Japan. \\ 
$^2$Yukawa Institute for Theoretical Physics, Kyoto University, 
Kyoto 606-8502, Japan. \\
$^3$Tata Institute of Fundamental Research, Homi Bhabha Road, 
Mumbai 400005, India.
}

\begin{abstract}
We explore the collision between two concentric spherical  thin shells. The inner shell is charged, whereas the outer one is either neutral or charged. In the situation we consider, the charge of the inner shell is larger than its gravitational mass, and the inside of it is empty and regular. Hence the domain just outside it is described by the overcharged Reissner-Nordstr\"{o}m geometry whereas the inside of it is Minkowski. First, the inner shell starts to shrink form infinity with finite kinetic energy, and then the outer shell starts to shrink from infinity with vanishing kinetic energy. The inner shell bounces on the potential wall and collides with the ingoing outer shell. The energy of collision between these shells at ``their center of mass frame" does not exceed the total energy of the system. By contrast, by virtue of the very large gamma factor of the relative velocity of the shells, the energy of collision between two of the constituent particles of these shells at their center of mass frame can be much larger than the Planck scale. This result suggests that the black hole or naked singularity is not necessary for ultra-high energy collision of particles.

\end{abstract}
\preprint{OCU-PHYS-379}
\preprint{AP-GR-104}
\preprint{YITP-12-101}


\maketitle

\section{Introduction}
Banados, Silk and West (BSW) recently pointed out that 
the Kerr black hole can act as a particle accelerator\cite{BSW}:  
Two particles dropped in from infinity
at rest, traveling along the timelike geodesics can collide with 
arbitrarily large energy at their center of mass frame, if the 
Kerr black hole is extremely spinning 
and orbital angular momentum of one of the particles takes a specific 
fine-tuned value. 
Then, the possible astrophysical implications of this process 
around the event horizon of the central supermassive black hole 
in the context of annihilations of the dark matter particles accreted
from the galactic halo
were investigated \cite{BSW2}. 
The BSW process of particle acceleration suffers from several 
drawbacks and limitations pointed out in \cite{Berti}:  
such as for example, a fine-tuning of the orbital angular momenta 
of the particles, a neglect of the self-gravity of particles in their analysis, 
and an upper bound on the spin of the Kerr black hole formed in our universe\cite{Thorne}. 
There were many investigations of this acceleration mechanism in the
background of Kerr as well as many other black holes\cite{BHaccel}. 

Two of the present authors, MP and PSJ (PJ), pointed out the other possibility.   
Particle collisions with arbitrarily large energy at the center of mass frame 
may occur in the naked singular Kerr spacetime if the total angular momentum 
of the central naked singularity is very close to  
the critical value\cite{Patil}. 
This mechanism is physically rather 
different from the BSW process. In the case of the 
BSW process, one of  
the two particles asymptotically approaches 
the event horizon by virtue of its 
special value of the orbital angular momentum. Since the event 
horizon is generated by the 
outward null geodesics,  the world line of the particle asymptotically becomes 
outward null. Hence, the relative velocity between this particle 
and another particle falling to the Kerr black hole 
can be very large. As a result, the center of mass energy of 
the collision between these particles can be arbitrarily large. 
By contrast, in the case of the PJ process, one of 
the two particles falls inward but eventually turns to outward 
due to the repulsive nature of the naked singularity 
or due to the centrifugal potential. 
Then, it can collide with another particle falling inward. 
In this mechanism, the large relative velocity between these 
two particles at the collision event 
can be achieved by virtue of the deep gravitational potential 
of the almost extreme naked singular geometry.

Like the BSW process, the PJ process of particle acceleration also 
would have certain drawbacks and limitations.  
As in the case of BSW's analysis, the self-gravities of the particles are neglected 
in PJ's analysis, and it is unclear whether 
the naked singular Kerr geometry appears in our universe\cite{Penrose}. 
However, in the case of the PJ process, no fine-tuning of the orbital 
angular momenta of the colliding particles is necessary.

The issue of the self-gravity of the point particles is 
difficult to deal with in general. 
If the effects of self-gravity and gravitational radiation 
are important in a 
collision between elementary particles, such a collision process 
must be described by the 
quantum gravity: The gravitational interaction between elementary 
particles is to be necessarily quantum in nature.
Hence, even if BSW or PJ analysis does not predict accurately the 
high energy collision of the elementary particles, their result
implies that collisions of elementary particles with the center 
of mass energy high enough to cause quantum gravitational 
interactions may occur in the Kerr spacetime. 
However, it is a fascinating primary question what happens 
when the center of mass energy of a collision becomes comparable 
to the total mass of the system within the framework of 
general relativity.

Unfortunately, it is very difficult to treat analytically the motion
of matter in the Kerr spacetime, 
if we will take into account the effect of their self-gravity. 
Hence, it is worthwhile to notice that the similar processes 
to both of the BSW and PJ processes may occur in the 
Reissner-Nordstr\"{o}m spacetime\cite{Zaslavskii,Nakao,Patil-2}: 
it is much easier than the case of the Kerr spacetime 
to treat analytically the motion of matter in the Reissner-Nordstr\"{o}m 
spacetime by virtue of its spherical symmetry.  
In the case of a process similar to the BSW one, a particle 
with charge of the same sign as that of the extreme 
Reissner-Nordstr\"{o}m black hole radially falls toward the 
black hole from infinity. 
If the charge of the particle is equal to its mass, 
the particle asymptotically approaches the event horizon. 
After the charged particle starts to fall, another neutral 
particle also starts a radial fall toward the 
black hole from infinity. Then, it  eventually overtakes and 
collides with the charged particle previously falling. 
The closer to the event horizon the collision event is, the 
higher will be its energy at their center of mass frame, and  
arbitrarily large collision energy is in principle possible. 
By contrast, in the case of the process similar to PJ's 
one, two radially moving neutral particles 
can collide with arbitrarily large energy at their center 
of mass frame. One of the two particles falls inward from infinity 
and eventually turns back outwards due to the repulsive nature 
of the central naked singularity. Then it collides with another 
particle which starts to fall towards the naked singularity 
after the first particle have started. 
The very large collision energy is possible due to the 
deep gravitational potential of almost extreme central naked 
singularity, like in the case of the naked singular 
Kerr spacetime.

A system of concentric spherical shells with infinitesimal width 
in the Reissner-Nordstr\"{o}m spacetime 
is very useful to study the effect of the self-gravity in the BSW and 
PJ processes, since their dynamical degrees of  freedom 
are finite and hence the system is tractable analytically. 
The stress-energy tensor diverges on the shells, 
since finite energy and momentum concentrate on the infinitesimally 
thin domains. This means 
that these shells are categorized into the so-called curvature 
polynomial singularity\cite{HE} through the Einstein equations. 
Since each shell has finite mass and momentum, the volume integral 
of the stress-energy tensor is finite: the components of the stress-energy tensor are 
distributional sources of Einstein's equations.  
Here we should note that the distributional source is a technical simplification 
usually adopted in theoretical study of gravitational physics. For example, in the framework of 
Newtonian gravity, massive point particles are useful idealization to study the celestial 
mechanics, although real stars have finite sizes.  By this idealization, the dynamical 
degrees of freedom of the system become finite, and as a result, the analysis is very easy. 
If the size of each star is much smaller than the size of the system, this point-particle 
approximation will give sufficiently accurate prediction about their orbits. 
The divergences of the gravitational potential and the gravitational force 
just at the point particle are artificial due to this technical idealization, and the gravity of each star 
is assumed to be so small that the Newtonian approximation is valid. 
By contrast, in the framework of 
general relativity, the point-particle approximation is impossible, since the point particle 
is so seriously singular that the spacetime metric cannot be defined on it.  
However the thin-shell approximation is possible even in the framework of general 
relativity, since the metric is defined on the infinitesimally thin shell 
even though the stress-energy tensor diverges on it. We can derive the equation of 
motion for the shell which is consistent with the Einstein equations 
by the so-called Israel formalism\cite{Israel,Poisson}. 
If the shell is highly symmetric (e.g. spherically symmetric), 
the degrees of freedom becomes finite, and as a result, 
the analysis is very easy. 
The divergence of the Ricci tensor 
at the shell is artificial due to this technical simplification. 
The thin-shell approximation is valid if the width of the shell is much smaller 
than the size of the system. 

Two of the present authors, MK, KN, and their collaborator, Tagoshi,  
studied the collision between two concentric spherical dust 
shells in the Reissner-Nordstr\"{o}m black hoke 
geometry; one of the two shells has a charge, whereas another 
is neutral. This example corresponds to the spherical shell 
version of the BSW process. 
By virtue of the spherical symmetry, they treated this system 
exactly and showed that the effect of the self-gravity makes  
the collision energy finite as long as we focus on the 
collision event observable for distant observers. 
Then, the present authors investigated the collision between two 
concentric neutral spherical dust shells in the 
naked singular Reissner-Nordstr\"{o}m geometry\cite{Patil-2}: 
This example corresponds to the spherical shell version of the PJ process. 
They showed that the upper bound on the energy of the collision between the shells appears due to the 
self-gravity of the shells also in this case 
as long as we focus on the collision event observable for the 
distant observers. But, the 
energy of the collision between two of the constituent particles 
of these shells can still exceed the Planck scale.  
Furthermore, in the case of the naked singular geometry, the timescale 
to occur the high-energy collision may be much shorter than the 
case that corresponds to the BSW process\cite{Patil-2}.

In this paper, we investigate
a collision between two concentric infinitesimally thin 
spherical shells made up of dust particles 
in a situation of no black hole and no naked singularity, taking 
into account their self-gravity exactly. The domain inside the inner 
shell is described by the Minkowskian 
geometry, whereas the outside of it is described by the over-charged 
Reissner-Nordstr\"{o}m geometry. The outer shell may or may not 
have charge, but the outside 
of it is assumed to be also described by the over-charged 
Reissner-Nordstr\"{o}m geometry. 
Even in this situation, very high energy collision between 
two of the constituent particles of these shells through a similar mechanism to 
the PJ process may occur, since the domain between these 
shells is described by the 
naked singular Reissner-Nordstr\"{o}m geometry. 

This paper is organized as follows. In Sec.~II, we briefly review Israel's 
formalism. Also in this section, we show the situation we consider 
and derive equations of motion for spherical shells.  In Sec.~III, 
the collision energy of two of constituent particles of the shells at their 
center of mass frame is shown. Sec. IV is devoted to summary and discussion. 

In this paper, we adopt the geometrized unit in which the speed of light and 
Newton's gravitational constant are unity.

\section{Equation of motion for spherical shells}
\begin{figure}
\begin{center}
\includegraphics[width=0.5\textwidth]{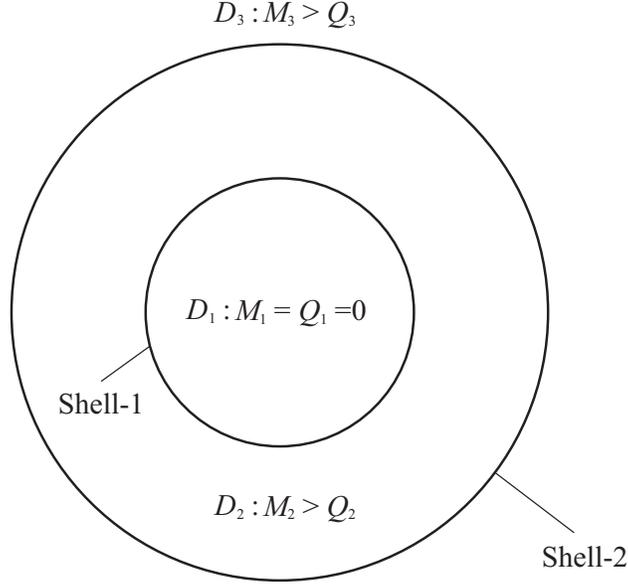}
\caption{\label{fig:shells}
This is a snapshot of the spherically symmetric spacetime divided into 
two domains $D_1$, $D_2$ and $D_3$ by the trajectories  
of the shell-1 and shell-2, i.e., $\Sigma_1$ and $\Sigma_2$.  
}
\end{center}
\end{figure}

We consider two concentric spherical shells which are infinitesimally thin. These 
shells may have $U(1)$ charge (see Fig.\ref{fig:shells}).

The trajectories of these shells in the spacetime are timelike hypersurfaces: the 
inner hypersurface is denoted by $\Sigma_1$, whereas the outer hypersurface 
is denoted by $\Sigma_2$. Correspondingly, the inner shell is called 
the shell-1, whereas 
the outer shell is called the shell-2. 
$\Sigma_1$ and $\Sigma_2$ divide the spacetime into 
three domains: the innermost domain is denoted by $D_1$, the middle one is 
denoted by $D_2$, and the outermost one is denoted by $D_3$. 
By the symmetry of this system, the geometry of the domain $D_i$ ($i=1,2,3$) 
is described by the Reissner-Nordstr\"{o}m solution whose 
line element is given by
\begin{eqnarray}
ds^2=-f_i (r)dt_i^2+\frac{1}{f_i(r)}dr^2
+r^2\left(d\theta^2+\sin^2\theta d\phi^2\right)  \label{RN}
\end{eqnarray}
with
\begin{equation}
f_i(r)=1-\frac{2M_i}{r}+\frac{Q_i^2}{r^2},   \nonumber
\end{equation}
where $M_i$ is the mass parameter and $Q_i$ is the $U(1)$ charge within the 
sphere of the radius $r$. We should note that the coordinate $t_i$ is 
not continuous across the shells, whereas $r$, $\theta$ and $\phi$ 
are everywhere continuous. The $U(1)$ gauge field in the domain $D_i$ is given by 
\begin{equation}
A_\mu=\frac{Q_i}{r}\delta^t_\mu.  \nonumber
\end{equation}

In the Reissner-Nordstr\"{o}m spacetime of the mass parameter $M$ and the charge $Q$, 
there is a spacetime singularity at 
$r=0$. This singularity is timelike and thus is necessarily locally naked.  
The location of the horizon in the Reissner-Nordstr\"{o}m spacetime 
is given by a positive root of the equation $f(r)\equiv1-2M/r+Q^2/r^2=0$ . 
There are two roots to this equation given by
\begin{equation}
r=r_{\pm}\equiv M \pm \sqrt{M^2-Q^2}.  \nonumber
\end{equation} 
There are two real positive roots of the equation 
if $M>Q$. The larger root $r=r_{+}$ 
is the location of the event horizon and this spacetime contains 
a spherically symmetric charged black hole. 
The smaller root $r=r_{-}$ corresponds to the Cauchy horizon associated with 
the timelike singularity at $r=0$.  
If $M=Q$, there is only one positive root. In this case, the black hole has 
a degenerate event horizon at $r=M=Q$; we call this the extreme black hole. 
In the case of $M<Q$, there is no real root of the equation $f(r)=0$. 
Thus, the event horizon is absent and the timelike 
singularity at $r=0$ is exposed 
to the asymptotic observer at infinity. This configuration 
thus contains a globally visible naked singularity. 

As mentioned, even though $\Sigma_A$ ($A=1,2$) are spacetime 
singularities, we can derive the equation of 
motion for each spherical shell which is consistent with the Einstein equations 
by the so-called Israel formalism\cite{Israel,Poisson}. 
Let us cover the neighborhood of one singular hypersurface $\Sigma_A$
by a Gaussian normal coordinate $\lambda$, 
where $\partial/\partial\lambda$ is perpendicular to $\Sigma_A$ and directs from 
$D_A$ to $D_{A+1}$. Then, the sufficient condition 
to apply Israel's formalism is that the stress-energy 
tensor is written in the form
\begin{equation}
T_{\mu\nu}=\sum_A S^{(A)}_{\mu\nu}\delta(\lambda-\lambda_A)  \nonumber
\end{equation}
where $\Sigma_A$ is located at $\lambda=\lambda_A$, 
$\delta(x)$ is Dirac's delta function, 
and $S^{(A)}_{\mu\nu}$ is finite and called the surface stress-energy tensor of $\Sigma_A$. 

The junction condition of the metric tensor is given as follows.  
We impose that the metric tensor $g_{\mu\nu}$ is continuous across $\Sigma_A$, 
but its derivative is not necessarily so. 
The unit normal vector to $\Sigma_A$ is $\partial/\partial \lambda$. 
Hereafter, we denote it 
by $n^\mu$. The intrinsic metric of $\Sigma_A$ is given by
\begin{equation}
h_{\mu\nu}=g_{\mu\nu}-n_\mu n_\nu.  \nonumber
\end{equation}
Then, the extrinsic curvature is defined by
\begin{equation}
K^{(i)}_{\mu\nu}=h^{\alpha}{}_\mu h^{\beta}{}_\nu \nabla^{(i)}_\alpha n_\beta,  \nonumber
\end{equation}
where $\nabla^{(i)}_\alpha$ is the covariant derivative 
with respect to the metric in the 
domain $D_i$. This extrinsic curvature describes how $\Sigma_A$ is embedded into 
the domain $D_i$. In accordance with Israel's formalism, the 
Einstein equations lead to
\begin{equation}
K^{(A+1)}_{\mu\nu}-K^{(A)}_{\mu\nu}=8\pi\left(S^{(A)}_{\mu\nu}
-\frac{1}{2}h_{\mu\nu}{\rm tr}S^{(A)}\right).
\label{j-con-0}
\end{equation}

In this article, the surface stress-energy tensors 
of the shells are assumed to be that of 
pressure-less matter, i.e., the dust
\begin{equation}
S^{(A)}_{\mu\nu}=\sigma_A u_\mu u_\nu,  \nonumber
\end{equation}
where $\sigma_A$ is the energy per unit area on $\Sigma_A$, and 
$u^\mu$ is the 4-velocity. We assume that $\sigma_A$ is positive. 
By the spherical symmetry, 
the motion of the shell-$A$ is described in the form $t_i=t_i(\tau)$ and 
$r=r(\tau)$, where $i=A$ or $i=A+1$, 
and $\tau$ is the proper time of the shell-$A$.  
The 4-velocity is given by
\begin{equation}
u^\mu=\left(\dot{t}_i,\dot{r},0,0\right),  \nonumber
\end{equation}
where a dot means a derivative with respect to $\tau$. 
Then, $n_\mu$ is given by
\begin{equation}
n_\mu=\left(-\dot{r},\dot{t}_i,0,0\right).  \nonumber
\end{equation}
Together with $u^\mu$ and $n^\mu$, the following unit vectors 
form an orthonormal frame;
\begin{eqnarray}
e_{(\theta)}^\mu&=&\left(0,0,\frac{1}{r},0\right),   \nonumber\\
e_{(\phi)}^\mu&=&\left(0,0,0,\frac{1}{r\sin\theta}\right).  \nonumber
\end{eqnarray}

The extrinsic curvature is obtained as
\begin{eqnarray}
K_{\mu\nu}^{(i)}u^\mu u^\nu&=&\frac{1}{f_{i}\dot{t}_i}\left(\ddot{r}+\frac{f'_i}{2}\right),   \nonumber\\
K^{(i)}_{\mu\nu}e_{(\theta)}^\mu e_{(\theta)}^\nu&=&
K^{(i)}_{\mu\nu}e_{(\phi)}^\mu e_{(\phi)}^\nu=-n^a\partial_a \ln r=-\frac{f_i}{r}\dot{t}_i   \nonumber
\end{eqnarray}
and the other components vanish, 
where a prime means a derivative with respect to $r$. 
By the normalization condition $u^\mu u_\mu=-1$, we have
\begin{equation}
\dot{t}_i=\pm\frac{1}{f_i}\sqrt{\dot{r}^2+f_i}~. \label{t-dot}
\end{equation}
We assume that there is no black hole initially, 
or equivalently, $f_i$ is initially everywhere positive. 
Then, $t_i$ corresponds to the time coordinate and its 
derivative $\dot{t}_i$ should be positive initially. 
Thus, we should choose the plus sign in the right hand side 
of Eq.~(\ref{t-dot}). 
If the shell-$A$ enters into a black hole and $f_i$ becomes negative, 
$t_{(i)}$ becomes spatial coordinate. If so, $\dot{t}_{(i)}$ may 
change its sign, and hence 
there is a possibility that we have to choose the minus sign 
in Eq.~\eq{t-dot}.  
But, as long as we do not mention, hereafter we assume the 
plus sign in Eq.~\eq{t-dot}. 

\subsection{The effective potential}

From $t$-$t$ and $\theta$-$\theta$ components of Eq.~\eq{j-con-0}, 
we obtain the following relation. 
\begin{eqnarray}
\sqrt{\dot{r}^2+f_{A+1}(r)}-\sqrt{\dot{r}^2+f_A(r)}
=-\frac{m_A}{r},
\label{j-con-1}
\end{eqnarray}
with
\begin{equation}
m_A:=4\pi\sigma_A r^2={\rm constant}.  \nonumber
\end{equation}
Note that $m_A$ is positive, since $\sigma_A$ is assumed to be positive. 

Let us rewrite Eq.~\eq{j-con-1} into the form of the energy equation. 
First, we write it in the form
\begin{equation}
\sqrt{\dot{r}^2+f_{A+1}(r)}=\sqrt{\dot{r}^2+f_A(r)}-\frac{m_A}{r},
\label{j-con-1-1}
\end{equation}
and then take a square of its both sides:
\begin{equation}
\dot{r}^2+f_{A+1}(r)=\dot{r}^2+f_A(r)+\left(\frac{m_A}{r}\right)^2-\frac{2m_A}{r}\sqrt{\dot{r}^2+f_A(r)}.
\label{j-con-1-2}
\end{equation}
Furthermore, we rewrite the above equation in the form
\begin{equation}
\sqrt{\dot{r}^2+f_A(r)}=\frac{r}{2m_A}\left[f_A(r)-f_{A+1}(r)+\left(\frac{m_A}{r}\right)^2\right].
\label{j-con-1-3}
\end{equation}
By taking a square of the both sides of the above equation, we have
\begin{equation}
\dot{r}^2+V_A(r)=0, 
\label{energy-eq}
\end{equation}
where
\begin{eqnarray}
V_A&\equiv&f_A(r)-\left(\frac{r}{2m_A}\right)^2 
\left[
f_A(r)-f_{A+1}(r)+\left(\frac{m_A}{r}\right)^2
\right]^2.
\label{potential}
\end{eqnarray}
The above equation is regarded as the energy equation for the shell-$A$. 
The function $V_A(r)$ corresponds to the effective potential. 
In the domain allowed 
for the motion of the shell-$A$, $V_A\leq0$ should be satisfied. 

Here note that the negativity of the right hand side of 
Eq.~\eq{j-con-1} implies to $f_{A+1}<f_A$ in 
the domain allowed for the motion of the shell-$A$. Furthermore, 
since the left hand side of 
Eq.~\eq{j-con-1-1} is non-negative, the right hand side of it 
should also be non-negative.  
Substituting Eq.~(\ref{j-con-1-3}) into the right hand side 
of Eq.~(\ref{j-con-1-1}), 
we find that the following inequality should be satisfied in the allowed domain; 
\begin{equation}
F_{A}(r)\equiv f_A(r)-f_{A+1}(r)-\left(\frac{m_A}{r}\right)^2
=\frac{2m_A{\cal E}_A}{r}-\frac{Q_{A+1}^2-Q_A^2+m_A^2}{r^2}
\geq 0, 
\label{seigen-0}
\end{equation}
where
\begin{equation}
{\cal E}_A\equiv\frac{M_{A+1}-M_A}{m_A}.  \nonumber
\end{equation}
We should note that ${\cal E}_A$ is not necessarily positive even in
the case of $m_A>0$, since
$M_{A+1}-M_A$ does not necessarily represent the energy of 
the shell-$A$ \cite{mass-parameters}.

\subsection{The allowed domain for the motion of the shells}

In the case of ${\cal E}_A>0$, Eq.~(\ref{seigen-0}) leads to
\begin{equation}
r\geq \rho_A
\label{seigen}
\end{equation}
where
\begin{equation}
\rho_A\equiv \frac{Q_{A+1}^2-Q_A^2+m_A^2}{2{\cal E}_Am_A}.   \nonumber
\end{equation}
In the case of ${\cal E}_A=0$, Eq.~(\ref{seigen-0}) leads to
\begin{equation}
Q_{A+1}^2-Q_A^2+m_A^2\leq0. \label{seigen-z0}
\end{equation}
In the case of ${\cal E}_A<0$, Eq.~(\ref{seigen-0}) leads to
\begin{equation}
r\leq \rho_A. \label{seigen-z1}
\end{equation}
Since $r$ is positive, the condition 
\begin{equation}
Q_{A+1}^2-Q_A^2+m_A^2< 0. \label{seigen-z2}
\end{equation}
should hold so that Eq.~(\ref{seigen-z1}) 
has a solution~\cite{innerblackhole}.

Here we should note that there is a possibility that if 
the shell-$A$ is in the domain 
with $Q_{A+1}>M_{A+1}$, there is a possibility that the domain with $V_A<0$ is not 
allowed for the motion of the shell-$A$. We see such a case 
in the next section.

In this article, we assume $M_1=0=Q_1$, $M_2<Q_2$ 
and $M_3 < Q_3$; the shell-1 is charged, whereas the shell-2 is 
not necessarily charged.    
The domain $D_1$ is then described by the Minkowski geometry, whereas 
$D_2$ and $D_3$ are described by the over-charged Reissner-Nordstr\"{o}m geometry. 
Hence $f_{A+1}$ ($A=1,2$) is everywhere positive.  
We can see that 
\begin{equation}
V_A(\rho_A)=f_{A+1}(\rho_A)>0.
\label{V-seigen}
\end{equation}
The above result implies that $r=\rho_A$ is necessarily in 
the domain forbidden for the motion of the shell-$A$.  
By the assumption on $M_i$ and $Q_i$($i=1,2,3$), since the left hand side of Eq.\eq{j-con-1-1} is positive, 
the function  $F_A(r)$ \eq{seigen-0} takes positive value.

\section{The energy of collision between two concentric shells}

In accordance with Ref. \cite{Nakao, Patil-2}, the energy 
$E_{\rm cm}$ of the collision between the shells 
in the ``center of mass frame" is given by
\begin{eqnarray}
E_{\rm cm}^2&=&-g_{\mu\nu}(m_1 u_1^\mu+m_2 u_2^\mu)
(m_1 u_1^\nu+m_2 u_2^\nu) \cr
&& \cr
&=&m_1^2+m_2^2 +\frac{2m_1m_2}{f_2} \left[
\sqrt{(f_2-V_1)(f_2-V_2)}+\sqrt{V_1V_2}
\right],
\label{shell-energy}
\end{eqnarray}
where we have assumed that the sign of $u_1^r$ is pooisite to that of $u_2^r$.
We are also interested in the energy $E_{\rm P}$ 
of the collision between two of constituent particles 
of these shells in their center of mass frame. We assume that all constituent 
particles  have an identical mass $m$. Then, we obtain
\begin{eqnarray}
E_{\rm P}^2&=&-g_{\mu\nu}m^2(u_1^\mu+u_2^\mu)
(u_1^\nu+u_2^\nu) \cr
&& \cr
&=&2m^2\left[1+
\frac{\sqrt{(f_2-V_1)(f_2-V_2)}+\sqrt{V_1V_2}}{f_2}
\right]. 
\label{cm-energy}
\end{eqnarray}

From the above equation, we can expect that $E_{\rm cm}$ and $E_{\rm P}$ are large if 
the collision occurs at the minimum of $f_2$. 
The function $f_2(r)$ takes a minimum value at 
\begin{equation}
r=r_{\rm min}\equiv\frac{Q_2^2}{M_2},  \nonumber
\end{equation}
and the minimum value $f_2(r_{\rm min})$ is equal to $1-(M_2/Q_2)^2$. 
If $Q_2$ is very close to $M_2$, the minimum value $f_2(r_{\rm min})$ is very small 
and $E_{\rm cm}$ and $E_{\rm P}$ may be very large. 
We shall estimate how large $E_{\rm cm}$ and $E_{\rm P}$ of the collision at $r=r_{\rm min}$ can be. 
For this purpose, we parametrize $Q_2$ as 
\begin{equation}
Q_2=(1+\epsilon)M_2,   \nonumber
\end{equation}
and we assume $0<\epsilon \ll 1$.

\subsection {The shell-1}
As mentioned, we assume both $M_1$ and $Q_1$ vanish, and hence  
we have $Q_2^2-Q_1^2+m_1^2=Q_2^2+m_1^2>0$. This implies 
that ${\cal E}_1$ should be positive, 
and, as a result, $M_2$ should also be positive: 
If ${\cal E}_1$ is negative, the areal radius $r$ should be less than 
zero by Eq.~(\ref{seigen-z1}), but this is not the case of our interest.  
Since we assume $Q_2>M_2$, $Q_2$ is necessarily positive.

The effective potential of the shell-1 is written in the form
\begin{equation}
V_1(r)=1-{\cal E}_1^2+\frac{{\cal E}_1}{m_1 r}(Q_2^2-m_1^2)
-\left(\frac{Q_2^2-m_1^2}{2m_1 r}\right)^2.
\label{V1}
\end{equation}
The roots of  $V_1(r)=0$ are 
\begin{equation}
r=\rho_\pm\equiv \frac{1}{2m_1({\cal E}_1\mp1)}\left(Q_2^2-\frac{M_2^2}{{\cal E}_1^2}\right). 
\label{t-point}
\end{equation} 
By some consideration, we see that  
the motion of the shell-1 is necessarily unbound ${\cal E}_1>1$ and its allowed domain 
is $r\geq\rho_+$, as long as we assume $Q_2>M_2$. The detail of the consideration is given 
in Appendix \ref{section:shell-1}. 

In order that the collision occurs at $r=r_{\rm min}$, $r=r_{\rm min}$ should be in 
a domain allowed for the motion of the shell-1. Thus, we have to impose 
$\rho_+<r_{\rm min}$. 
Together with the condition $M_2={\cal E}_1m_1>m_1$, 
the condition $\rho_+<r_{\rm min}$ leads to 
\begin{equation}
{\cal E}_1 >1+\sqrt{1-\left(\frac{M_2}{Q_2}\right)^2} =1+(2\epsilon)^{1/2}+{\cal O}(\epsilon).  \nonumber
\end{equation}
If ${\cal E}_1$ is slightly larger than unity, the shell-1 shrinks to the radius less than $r_{\rm min}$. 

The effective potential of the shell-1 at $r=r_{\rm min}$ is given by
\begin{eqnarray}
V_1(r_{\rm min})&=&
1-\frac{1}{4}\left[{\cal E}_1+\frac{1}{(1+\epsilon)^2{\cal E}_1}\right]^2 \cr
&&\cr
&=&-\frac{1}{4}\left({\cal E}_1-\frac{1}{{\cal E}_1}\right)^2
+\left(1+\frac{1}{{\cal E}_1^2}\right)\epsilon+{\cal O}(\epsilon^2). 
\label{V1-rmin}
\end{eqnarray}
If the shell-1 is almost marginally bound, 
i.e., ${\cal E}_1=1+(2\epsilon)^{1/2}\alpha$ with 
$\alpha$ larger than one but of order unity, we have
\begin{equation}
V_1(r_{\rm min})=-2(\alpha^2-1)\epsilon+{\cal O}(\epsilon^2).  \nonumber
\end{equation}
From the above result, we can see that, 
in the almost marginally bound case, the speed of the 
shell-1 at $r=r_{\rm min}$, which is given 
by $\sqrt{-V_1(r_{\rm min})}$, 
is very small. The high speed of the shell-1 at $r=r_{\rm min}$ is achieved  
only if ${\cal E}_1-1$ is almost equal to or larger than unity.

\subsection{The shell-2}
We assume that the shell-2 starts to fall inward from infinity 
at rest. This assumption is equivalent to 
the condition ${\cal E}_2=1$, i.e., $M_3-M_2=m_2$. As long as we see
the effective potential $V_2$,
${\cal E}_2=-1$ seems to also be a solution, but it is not true due
to Eq.~(\ref{seigen-z1}):
negative ${\cal E}_2$ implies the bound motion. 

We adopt the following parametrization;
\begin{eqnarray}
M_3&=&M_2+\mu M_2=(1+\mu)M_2,   \nonumber\\
Q_3&=&(1+q)Q_2=(1+\epsilon)(1+q)M_2,   \nonumber\\
m_2&=&\mu M_2.  \nonumber
\end{eqnarray}
Note that $\mu$ is positive since $m_2$ is positive. 

By the careful analysis, we see that, in order that the shell-2 which is at rest at infinity  
starts to fall inward and collides with the shell-1 at 
$r=r_{\rm min}$, the two cases appear.  
In the first case, the following condition holds: 
\begin{eqnarray}
0<\mu<q(1+\epsilon)+\epsilon \label{shell-2_con-2}
\end{eqnarray}
for
\begin{eqnarray}
-\frac{\epsilon}{1+\epsilon} < q \leq 0. \label{shell-2_con-1}
\end{eqnarray}
By contrast, in the second case, the following condition holds: 
\begin{eqnarray}
\mu_{\rm m}<\mu<q(1+\epsilon)+\epsilon, \label{shell-2_con-4} 
\end{eqnarray}
for
\begin{eqnarray}
0<q< q_*, \label{shell-2_con-3} 
\end{eqnarray}
where
\begin{eqnarray}
\mu_{\rm m}&=&(1+\epsilon)\left\{1+\epsilon-\sqrt{(1+\epsilon)^2-1}\right\}
\left[
\sqrt{1+\frac{q(q+2)}{\left\{1+\epsilon-\sqrt{(1+\epsilon)^2-1}\right\}^2}}-1
\right], \label{mu-bound}\\
&&\cr
q_*&=&\frac{\epsilon^2+\sqrt{\epsilon(1+\epsilon)(2+\epsilon+\epsilon^2)}}{2(1+\epsilon)}. 
\label{q-star}
\end{eqnarray}
The derivation of the above conditions is given in Appendix \ref{section:shell-2}.

\subsubsection{The case with negative charge or without charge}
From the conditions (\ref{shell-2_con-1}), we obtain 
\begin{equation}
q=-\frac{\epsilon w_-}{1+\epsilon}~~~~~~~{\rm with}~~0\leq w_-<1. \label{para-q-1}
\end{equation}
Substituting the above expression into Eq.~(\ref{shell-2_con-2}), we have
\begin{equation}
0<\mu < \epsilon (1-w_-).  \nonumber
\end{equation}
Thus we get
\begin{equation}
\mu=\epsilon(1-w_-)x_-~~~~~~~~~{\rm with}~~0<x_-<1. \label{para-beta-1}
\end{equation}

By using the expressions (\ref{para-q-1}) and (\ref{para-beta-1}), we have
\begin{equation}
V_2(r_{\rm min})=-\frac{\left[(1-w_-)x_-+w_-\right]^2}{(1-w_-)^2x_-^2}+{\cal O}(\epsilon)<0.
\label{V2-rmin(-)}
\end{equation}

\subsubsection{The case with positive charge}

From the conditions (\ref{shell-2_con-3}), we obtain 
\begin{equation}
q=\left(\frac{\epsilon}{2}\right)^{1/2}w_+~~~~~{\rm with}
~~0 \leq w_+<\epsilon^{-1/2}\sqrt{2}q_*.    \label{p-para-q-1}
\end{equation}
Note that $|\epsilon^{-1/2}\sqrt{2}q_*-1| \ll1$. 
From Eq.~(\ref{shell-2_con-4}), we have
\begin{equation}
\mu=\mu_{\rm m}(1-x_+)+\left[q(1+\epsilon)+\epsilon\right]x_+~~~~{\rm with}~~0<x_+<1.  \nonumber
\end{equation}
Substituting Eq.~(\ref{p-para-q-1}) into the above expression, we have
\begin{equation}
\mu=\left(\frac{\epsilon}{2}\right)^{1/2}w_+ 
+\epsilon \left(w_+-w_+x_++x_+\right)
+{\cal O}(\epsilon^{3/2}). \label{p-para-beta-1}
\end{equation}
By using the expressions (\ref{p-para-q-1}) and (\ref{p-para-beta-1}), we have
\begin{equation}
V_2(r_{\rm min})=-\frac{2\epsilon}{ w_+^2}x_+(1-w_+)\left\{2w_+ + x_+(1-w_+)\right\}
+{\cal O}(\epsilon^{3/2})<0.
\label{V2-rmin(+)}
\end{equation}

\subsection{The energy of collision}

We consider the collision of two shells such that
the shell-1, which is initially ingoing, turns back as an
outgoing shell and then collides with the ingoing shell-2 at $r = r_{\rm min}$.
In this subsection, we estimate the center of mass energy for this collision.

\subsubsection{The case with negative charge or without charge}

By using Eqs.~(\ref{shell-energy}), (\ref{V1-rmin}) and (\ref{V2-rmin(-)}), we have
\begin{eqnarray}
E_{\rm cm}&=&\frac{M_3}{{\cal E}_1}\left[1+({\cal E}_1^2-1)\left\{(1-w_-)x_-+w_-\right\}\right]^{1/2}
+{\cal O}(\epsilon) .  \nonumber
\end{eqnarray}
The above equation implies that the collision energy 
of the shells at their center of mass frame  
does not exceed the total energy of the system $M_3$. 
However the gamma factor of the relative velocity between the shells can be very large:  
\begin{equation}
\Gamma\equiv -g_{\mu\nu}u_1^\mu u_2^\nu=\frac{1}{2\epsilon} 
\left(\frac{(1-w_-)x_-+w_-}{(1-w_-)x_-}\right)
\left(\frac{{\cal E}_1^2-1}{{\cal E}_1}\right)+{\cal O}(\epsilon^0),   \nonumber
\end{equation}
and hence $E_{\rm P}$ can also be very large. 
By using Eqs.~(\ref{cm-energy}), (\ref{V1-rmin}) and (\ref{V2-rmin(-)}), we have
\begin{eqnarray}
E_{\rm P}&=&\frac{m}{\epsilon^{1/2}}\left(\frac{(1-w_-)x_-+w_-}{(1-w_-)x_-}\right)^{1/2}
\left(\frac{{\cal E}_1^2-1}{{\cal E}_1}\right)^{1/2}+{\cal O}(\epsilon^{1/2}) \cr
&& \cr
&\simeq& 9.4\times10^{18}
\left(
\frac{(1-w_-)x_-+w_-}{(1-w_-)x_-}
\right)^{1/2}
\left(\frac{{\cal E}_1^2-1}{{\cal E}_1}\right)^{1/2}
\left(\frac{m}{m_{\rm p}}\right)
\left(\frac{10^{-38}}{\epsilon}\right)^{1/2}{\rm GeV},
\label{Ecm-n-1(-)}
\end{eqnarray}
where $m_{\rm p}$ is the mass of a proton. 

The above result implies that $E_{\rm P}$ can be indefinitely large
in the limit of
 $(1-w_-)x_-\rightarrow0$ with $[(1-w_-)x_-+w_-]^{1/2}$ fixed. However, we
should note that the number of constituent
 particles of the shell $N\equiv m_2/m$ should be 
much larger than unity so that the continuum approximation is valid.  
By the definition of $N$, we have 
\begin{equation}
N=\epsilon(1-w)x\frac{M_2}{m}=1.2\times10^{19}(1-w_-)x_-\left(\frac{\epsilon}{10^{-38}}\right)
\left(\frac{M_2}{M_\odot}\right)\left(\frac{m_{\rm p}}{m}\right),  \nonumber
\end{equation}
where $M_\odot$ is the solar mass. 
The above result implies that $(1-w_-)x_-$ must not be too small so that
$N$ is larger than unity.
But our result still implies that, through the collision between two shells
with charge of the different sign or with no charge,
the energy of the collision between their constituent particles 
can exceed the Planck scale.  

\subsubsection{The case with positive charge}

By using Eqs.~(\ref{shell-energy}), (\ref{V1-rmin}) and (\ref{V2-rmin(+)}), we have
\begin{eqnarray}
E_{\rm cm}&=&\frac{M_3}{{\cal E}_1}\left[1+\frac{1}{2}
({\cal E}_1^2-1)\left\{\sqrt{w_+^2+V(w_+,x_+)}+\sqrt{V(w_+,x_+)}\right\}\right]^{1/2}
+{\cal O}(\epsilon^{1/2}) ,  \nonumber
\end{eqnarray}
where
\begin{equation}
V(w,x):=x(1-w)[2w+x(1-w)].
\end{equation}
The above equation implies that the collision energy 
of the shells at their center of mass frame  
does not exceed the total energy of the system $M_3$ also in the positive charge case. 
However, also in this case, 
since the gamma factor of the relative velocity between the shells can be very large, 
$E_{\rm P}$ can also be very large. 
By using Eqs.~(\ref{cm-energy}), (\ref{V1-rmin}) and (\ref{V2-rmin(+)}), we have
\begin{eqnarray}
E_{\rm P}&=&\frac{m}{(2\epsilon)^{1/4}}
\left[
\left({\cal E}_1-\frac{1}{{\cal E}_1}\right)
\frac{\sqrt{w_+^2+V(w_+,x_+)}+\sqrt{V(w_+,x_+)}}{w_+}
\right]^{1/2}
+{\cal O}(\epsilon^{1/4}) \cr
&& \cr
&\simeq& 7.9\times10^{18}
\left[
\left({\cal E}_1-\frac{1}{{\cal E}_1}\right)
\frac{\sqrt{w_+^2+V(w_+,x_+)}+\sqrt{V(w_+,x_+)}}{w_+}
\right]^{1/2} \cr
&&\cr
&\times&\left(\frac{m}{m_{\rm p}}\right)
\left(\frac{10^{-76}}{\epsilon}\right)^{1/4}{\rm GeV}.
\label{Ecm-n-1(+)}
\end{eqnarray}
In this case, the fine tuning of $\epsilon$ required for the ultra-high energy collision 
is severer than the negative charge or neutral case.  

The above result implies that $E_{\rm P}$ can be indefinitely large
in the limit of 
 $w_+ \rightarrow0$ with $x_+$ fixed. However, we
should note that we have assumed that $\epsilon^{1/2}w_+ \gg \epsilon x_+$, i.e., 
$w_+\gg \epsilon^{1/2}x_+$,  
when we derived Eq.~(\ref{p-para-beta-1}). 
Thus, in Eq.~(\ref{Ecm-n-1(+)}), such a limit must not be taken. 
But our result still implies that, through the collision between two shells
with the same sign of charge, 
the energy of the collision between their constituent particles 
can exceed the Planck scale.

\section{Summary and discussion}

We studied a collision between two concentric spherical 
thin dust shells: the inner shell 
is over-charged, whereas the outer one may or may not be charged. 
The domain in the inner shell is assumed to be described by 
the Minkowskian geometry, whereas 
the domain between the two shells and the domain outside the outer shell are 
assumed to be described by the over-charged Reissner-Nordstr\"{o}m geometry. 
First, the inner shell starts to fall inwards from infinity with 
finite initial velocity and eventually 
turns outward due to its self-electric force which overcomes 
its self-gravitational force. 
The outer shell starts to fall inward from infinity with 
vanishing initial velocity after a sufficient time after the inner shell has left. 
It will collide with inner shell going outward. 

We found that if the Reissner-Nordstr\"{o}m geometry between these shells is almost extreme 
but a bit over-charged, the energy of collision between two of 
constituent particles of these shells at their center of mass frame can be much 
larger than the Planck scale even if the mass of particle is order of proton mass, 
whereas the energy of collision between two shells at their center of mass frame can not 
exceed the total energy of the system.  
We would like to stress that 
neither black hole horizon nor naked singularity is necessary 
to achieve very high energy of the collision between constituent particles of the shells 
in this case. The necessary 
condition to achieve the high energy is that the collision occurs at 
$r=r_{\rm min}\equiv Q_2^2/M_2$, where $Q_2$ and $M_2$ are the charge  
and mass parameters of the Reissner-Nordstr\"{o}m geometry between these shells.

It is important fact that the inner over-charged shell can 
shrink to $r<r_{\rm min}$ if the 
initial inward velocity of the inner shell at infinity exceed a small threshold value.  
Furthermore, the inner shell can shrink to an arbitrarily small radius: 
we can see from Eq.~(\ref{t-point}) that, in the limit ${\cal E}_1\rightarrow\infty$, 
the radius of the turning point $r=\rho_+$ vanishes. This means that 
even if no naked singularity forms, a domain described by 
the over-charged Reissner-Nordstr\"{o}m geometry arbitrarily 
close to the naked singularity may appear as a transient phenomenon. 
The physical phenomena  similar to the PJ process may occur in such a domain, 
even though no naked singularity forms. 

As mentioned, the infinitesimally thin shell is a technical simplification. 
Although the stress-energy tensor of the shell diverges, this prescription is based on 
the assumption that the self-gravity of each constituent particle of the thin shells 
is negligible: the local dynamics 
of  each particle is described in the framework of the special relativity. 
However, if the collision energy of the particles at the center of mass frame exceeds the Planck scale, 
the colliding particles will make the spacetime highly curved, and black holes may form. 
Their motions after ultra-high energy collision 
cannot be described in the framework of the special relativity.  
This means that the motion of the shells after the collision is highly non-trivial problem, 
but this issue is out of the scope of this article.

\section*{Acknowledgments}
MK is supported by the JSPS Grant-in-Aid for Scientific Research
No.23$\cdot$2182.
KN thanks H. Ishihara and colleagues in the group of elementary
particle physics and gravity
at Osaka City University for useful discussions. 

\appendix
\section{The shell-1}
\label{section:shell-1}

Here, we show that ${\cal E}_1$ must be larger than unity and 
the allowed domain for the motion of the shell-1 is restricted to $r\geq\rho_+$, where 
$\rho_+$ is given in Eq.~(\ref{t-point}). 
As mentioned, ${\cal E}_1$ should be positive, and an inequality $0<M_2<Q_2$ holds. 

First, we note that there is only one root of the equation $dV_1(r)/dr=0$: the root is given by
\begin{equation}
r=\rho_{\rm m}\equiv
\frac{Q_2^2-m_1^2}{2{\cal E}_1m_1}
=\frac{1}{2{\cal E}_1m_1}\left(Q_2^2-\frac{M_2^2}{{\cal E}_1^2}\right).
\nonumber
\end{equation}
We can see that $\rho_{\rm m}$ is positive if and only 
if ${\cal E}_1$ is larger than $M_2/Q_2$ which is 
less than unity. 
Hence, in the case of ${\cal E}_1>M_2/Q_2$, 
$V_1(r)$ has one extremum at $r=\rho_{\rm m}$. We can easily see 
that this extremum is the maximum 
and $V_1(\rho_{\rm m})=1$. 
Here note that $V_1(r)\rightarrow 1-{\cal E}_1^2$ in the 
limit $r\rightarrow \infty$. 
Hence, in the case of $0<{\cal E}_1<1$, the spatial asymptotic 
region is a domain forbidden 
for the motion of the shell-1. 
Since $V_1(r)$ is monotonically decreasing in the domain 
of $r>\rho_{\rm m}$ in the case of 
${\cal E}_1=1$, we can see $V_1(r)>0$ in the domain of $r\geq\rho_{\rm m}$. 
Thus the spatial asymptotic region is the forbidden domain 
also in the case of ${\cal E}_1=1$.  
Only in the case of ${\cal E}_1>1$ the spatial asymptotic 
domain is allowed for the 
motion of the shell-1.

\subsection{$0<{\cal E}_1<M_2/Q_2$}
In this case, $V_1(r)$ is a monotonically increasing function of $r$, and  
only $r=\rho_+$ is a positive root of $V_1(r)=0$. 
$V_1(r)\geq0$ in the domain $r\geq \rho_+$, whereas $V_1(r)<0$ 
in the domain of $r <\rho_+$. 
Since $V_1(\rho_1)>0$ by Eq.~(\ref{V-seigen}), 
we find that $\rho_1>\rho_+$, and hence even if $V_1(r)\leq0$ in the domain  
$r\leq\rho_+$, the domain $r\leq\rho_+$ cannot be allowed for 
the motion of the shell-1 by Eq.~(\ref{seigen}).  
As a result, in the case of $0<{\cal E}_1<M_2/Q_2$, there is no domain allowed for the 
motion of the shell-1.

\subsection{${\cal E}_1=M_2/Q_2$} 
There is no allowed domain for the motion of the shell-1,  
since $V_1$ is identically equal to $1-(M_2/Q_2)^2$ 
which is positive.

\subsection{$M_2/Q_2<{\cal E}_1\leq1$}
Only $r=\rho_-$ is a positive root of $V_1(r)=0$. 
Since $V_1(r)\geq0$ in the domain $r\geq \rho_-$, whereas 
$V_1(r)<0$ in the domain $r <\rho_-$,  
the situation of this case is very similar to the case of $0<{\cal E}_1<M_2/Q_2$. 
By the similar argument to the case of $0<{\cal E}_1<M_2/Q_2$, 
we can see that there is no allowed domain for the motion of 
the shell-1 in the case of $M_2/Q_2<{\cal E}_1\leq1$.

\subsection{${\cal E}_1>1$}
Both $r=\rho_+$ and $r=\rho_-$ are positive roots of $V_1(r)=0$. 
In this case, $V_1(r)\leq0$ in the domain $r\leq\rho_-$ or $r\geq\rho_+$, 
whereas $V_1(r)>0$ in the domain $\rho_-<r<\rho_+$. 
We find from Eq.~(\ref{V-seigen}) that $\rho_-<\rho_1<\rho_+$ holds, 
and hence the domain $r<\rho_-$ 
is forbidden for the motion of the shell-1 by Eq.~(\ref{seigen}). 
The allowed domain for the motion of the shell-1 is only $r\geq\rho_+$.  

\vskip0.5cm
In summary, the shell-1 is necessarily unbound 
${\cal E}_1>1$ and the allowed domain 
is $r\geq\rho_+$, as long as we assume $0<M_2<Q_2$.

\section{The shell-2}
\label{section:shell-2}

In this appendix, we derive the conditions (\ref{shell-2_con-2}) -- (\ref{q-star}). 

\subsection{From the over-charge condition}

Since we assume that the domain $D_3$ is described by the 
over-charged Reissner-Nordstr\"{o}m geometry $Q_3>M_3$, we have
\begin{equation}
\mu<(1+\epsilon)q+\epsilon.  \label{denka-seigen-1}
\end{equation}
Since $\mu$ is positive, $(1+\epsilon)q+\epsilon$ should be positive
so that the above
inequality has a solution. Hence, we have
\begin{equation}
q>-\frac{\epsilon}{1+\epsilon}. \label{denka-seigen-2}
\end{equation}

\subsection{From the marginally-bound condition}

In the marginally bound case, the effective potential $V_2$ of the shell-2 is written in the form
\begin{eqnarray}
V_2(r)=-\frac{M_2}{\mu r}A_1(\mu;q) -\left(\frac{M_2}{2\mu r}\right)^2A_2(\mu;q), \nonumber
\end{eqnarray}
where
\begin{eqnarray}
A_1(\mu;q)&\equiv& \mu(\mu+2)-(1+\epsilon)^2q(q+2), \cr
A_2(\mu;q)&\equiv&\left[(1+\epsilon)q-\mu\right]\left[(1+\epsilon)q+\mu\right]
\left[(1+\epsilon)(q+2)-\mu\right] \left[(1+\epsilon)(q+2)+\mu\right]. \nonumber
\end{eqnarray}
Since $V_2(r)$ should be negative in the domain allowed 
for the motion of the shell-2, the following condition should be 
satisfied so that the shell-2 can start to fall inward from infinity;
\begin{equation}
A_1(\mu;q)>0
\label{MB-seigen-1}
\end{equation}
is satisfied, or both $A_1=0$ and $A_2>0$ hold simultaneously.

From Eq.~(\ref{MB-seigen-1}), we obtain the following constraint on $\mu$;  
in the case of $q(q+2)>0$, 
\begin{equation}
\mu>b  \label{MB-seigen-2}
\end{equation}
should hold, where $b$ is a larger root of $A_1(b;q)=0$: 
\begin{equation}
b=-1+ \sqrt{1+(1+\epsilon)^2q(q+2)}, \nonumber
\end{equation}
whereas, in the case of $q(q+2)\leq0$, 
Eq.~(\ref{MB-seigen-1}) necessarily holds, since $\mu$ is positive. 

The solution of the inequality $q(q+2)\leq0$ is $-2\leq q \leq 0$. However, 
since $-\epsilon(1+\epsilon)^{-1}$ is larger than $-2$ for
$\epsilon>0$, we should consider from
 Eq.~(\ref{denka-seigen-2}) that 
the solution of the inequality $q(q+2)\leq0$ is
$-\epsilon(1+\epsilon)^{-1}<q\leq0$.

The solution of the inequality $q(q+2)>0$ is $q<-2$ or $0<q$.  
However, Eq.~(\ref{denka-seigen-2}) implies that 
$q$ must be larger than $-2$, and hence we should  consider that 
the solution of the inequality  $q(q+2)>0$ is $q>0$. 
In the case of $q(q+2)>0$, $b$ must be less than 
$(1+\epsilon)q+\epsilon$ so that 
Eqs.~(\ref{denka-seigen-1}) and (\ref{MB-seigen-2}) 
are consistent with each other.  
By the definition of $b$, we have
\begin{equation}
 (1+\epsilon)q+\epsilon-b=
 (1+\epsilon)(1+q)-\sqrt{(1+\epsilon)^2(1+q)^2-\epsilon(2+\epsilon)}. \nonumber
\end{equation}
We can see from the above equation that $(1+\epsilon)q+\epsilon-b>0$ 
is satisfied because of $q>0$.  Hence, the consistency between 
Eqs.~(\ref{denka-seigen-1}) and (\ref{MB-seigen-2}) adds no further constraint.

In the case of $q(q+2)>0$, i.e., $q>0$,  
$A_1$ vanishes  only for $\mu=b$.  In this case, we can see 
\begin{eqnarray}
q(1+\epsilon)-b &<&0, \label{ineq-1}\\
q(1+\epsilon)+b &>&0, \label{ineq-2}\\
(q+2)(1+\epsilon)-b &>&0, \label{ineq-3}\\
(q+2)(1+\epsilon)+b &>&0. \label{ineq-4}
\end{eqnarray}

First, we prove Eq.~(\ref{ineq-1}). We have
\begin{equation}
q(1+\epsilon)-b=q(1+\epsilon)+1-\sqrt{1+(1+\epsilon)^2q(q+2)}. \nonumber
\end{equation}
Since $q>0$, we can see
\begin{equation}
\left[q(1+\epsilon)+1\right]^2-\left[1+(1+\epsilon)^2q(q+2)\right]
=-2\epsilon(1+\epsilon)q<0. \nonumber
\end{equation}
Hence, we obtain Eq.~(\ref{ineq-1}). 

Next, we prove Eq.~(\ref{ineq-2}). We have
\begin{equation}
q(1+\epsilon)+b=q(1+\epsilon)-1+\sqrt{1+(1+\epsilon)^2q(q+2)}. \nonumber
\end{equation}
Since $q>0$, we have
\begin{equation}
\left[q(1+\epsilon)-1\right]^2-\left[1+(1+\epsilon)^2q(q+2)\right]
=-2(2+\epsilon)(1+\epsilon)q<0. \nonumber
\end{equation}
Hence, we have Eq.~(\ref{ineq-2}). 

By the over-charge condition (\ref{denka-seigen-1}), i.e., 
$q(1+\epsilon)+\mu>-\epsilon$, we can see that 
Eqs.~(\ref{ineq-3}) and (\ref{ineq-4}) should be satisfied; 
\begin{eqnarray}
(q+2)(1+\epsilon)-b&=&q(1+\epsilon)-b+2(1+\epsilon) >-\epsilon+2(1+\epsilon)=2+\epsilon>0, \cr
(q+2)(1+\epsilon)+b&=&q(1+\epsilon)-b+2b >2+\epsilon+2b>0. \nonumber
\end{eqnarray}

Equations (\ref{ineq-1}) - (\ref{ineq-4}) 
imply that $A_2$ is negative, and thus this case is not of our interest. 

In summary, we have found that the over-charge condition (\ref{denka-seigen-1}) 
and marginally bound 
conditions (\ref{MB-seigen-1}) and (\ref{MB-seigen-2}) imply  
\begin{eqnarray} 
&&0<\mu <(1+\epsilon)q +\epsilon~~~~{\rm for}~~ -\frac{\epsilon}{1+\epsilon} < q \leq0, \label{result-m0}\\ 
&&b<\mu<(1+\epsilon)q+\epsilon~~~{\rm for}~~~~~0<q. \label{result-p0}
\end{eqnarray}

\subsection{To reach $r=r_{\rm min}$}

In order that a collision can occur at $r=r_{\rm min}$, we have
\begin{equation}
F_2(r_{\rm min})=-(1+\epsilon)^{-4}G(\mu;q) > 0 \label{F-seigen} 
\end{equation}
and
\begin{equation}
V_2(r_{\rm min})=-\frac{1}{4\mu^2}(1+\epsilon)^{-4}H(\mu;q) \le 0, 
\label{V2-seigen}
\end{equation}
where
\begin{eqnarray}
G(\mu;q)&=&\mu^2-2(1+\epsilon)^2\mu+(1+\epsilon)^2q(q+2), \\
&&\cr
H(\mu;q)&=&\left[\mu^2+2(1+\epsilon)^2\mu-(1+\epsilon)^2q(q+2)\right]^2
-4\epsilon(2+\epsilon)(1+\epsilon)^2\mu^2.
\end{eqnarray}

First, we consider the condition (\ref{F-seigen}). 
The condition (\ref{F-seigen}) is rewritten in the form of $G(\mu;q) <  0$. 
It can be easily seen that 
$q(q+2)$ should be less than $(1+\epsilon)^2$ so that this 
inequality has a solution with respect to $\mu$. 
If $q(q+2)\leq0$, i.e., $-\epsilon(1+\epsilon)^{-1}<q\leq0$ holds, 
this constraint is satisfied. On the other hand, 
if $q(q+2)>0$, i.e., $q>0$ holds, the constraint $q(q+2)<(1+\epsilon)^2$ leads to
\begin{equation}
0<q<\sqrt{1+(1+\epsilon)^2}-1. \nonumber
\end{equation} 
Then, we obtain
\begin{eqnarray}
&&0<\mu<b_+~~~~~~~{\rm
 for}~~~~ -\frac{\epsilon}{1+\epsilon} < q \leq 0, \label{result-m1}\\
&&b_-<\mu<b_+~~~~~{\rm for}~~~~0<q< \sqrt{1+(1+\epsilon)^2}-1, \label{result-p1}
\end{eqnarray}
where $b_\pm$ are the roots of $G(b_\pm;q)=0$: 
\begin{equation}
b_\pm=(1+\epsilon)^2\pm(1+\epsilon)\sqrt{(1+\epsilon)^2-q(q+2)}. \nonumber
\end{equation}

In order to see which are sharp conditions, 
the pair of Eqs.~(\ref{result-m0}) and (\ref{result-p0}) or 
the pair of Eqs.~(\ref{result-m1}) and (\ref{result-p1}), 
we investigate which is larger, $b_+$ or $q(1+\epsilon)+\epsilon$. 
We have
\begin{equation}
q(1+\epsilon)+\epsilon-b_+=
q(1+\epsilon)-(1+\epsilon+\epsilon^2)-(1+\epsilon)\sqrt{(1+\epsilon)^2-q(q+2)}. \nonumber
\end{equation}
Here we define the following quantity
\begin{equation}
c\equiv q(1+\epsilon)-(1+\epsilon+\epsilon^2). \nonumber
\end{equation}
If $c$ is negative, the quantity $q(1+\epsilon)+\epsilon-b_+$ is also negative. 
We can easily see that $c$ is negative in the case of
$-\epsilon(1+\epsilon)^{-1} < q\leq 0$.
 By contrast, in the case of $0<q<\sqrt{1+(1+\epsilon)^2}-1$, we have 
\begin{eqnarray}
c&<&\left[\sqrt{1+(1+\epsilon)^2}-1\right](1+\epsilon)-(1+\epsilon+\epsilon^2) \cr
&=&(1+\epsilon)\sqrt{1+(1+\epsilon)^2}-\left[1+(1+\epsilon)^2\right]. \nonumber
\end{eqnarray}
We can see
\begin{equation}
\left[(1+\epsilon)\sqrt{1+(1+\epsilon)^2}\right]^2-\left[1+(1+\epsilon)^2\right]^2
=-\left[1+(1+\epsilon)^2\right]<0. \nonumber
\end{equation}
Hence, $c$ is negative also in the case of $0<q<\sqrt{1+(1+\epsilon)^2}-1$. 
Thus, we obtain 
\begin{equation}
q(1+\epsilon)+\epsilon<b_+~~~~{\rm for}~~
-\frac{\epsilon}{1+\epsilon} < q< \sqrt{1+(1+\epsilon)^2}-1. \nonumber
\end{equation}

Now, from the over-charge condition (\ref{denka-seigen-1}), the marginally-bound 
condition (\ref{MB-seigen-1}) and (\ref{MB-seigen-2}), the constraint
(\ref{F-seigen}),
 we have
 \begin{eqnarray}
&&0<\mu<q(1+\epsilon)+\epsilon~~~~~~~~~~~~~~~~~~{\rm for}~~~~ 
-\frac{\epsilon}{1+\epsilon} < q \leq 0, \label{ineq-5}\\
&&{\rm max}[b_-,b]<\mu<q(1+\epsilon)+\epsilon~~~~~~~{\rm for}~~~~0<q< \sqrt{1+(1+\epsilon)^2}-1.
\label{ineq-6}
\end{eqnarray}

Here, let us focus on the condition (\ref{ineq-6}). 
In order that Eq.~(\ref{ineq-6}) is satisfied, $b_-$ 
must be less than $q(1+\epsilon)+\epsilon$. 
Thus, we have
\begin{equation}
q(1+\epsilon)+\epsilon-b_-=c+(1+\epsilon)\sqrt{(1+\epsilon)^2-q(q+2)}>0. \nonumber
\end{equation}
Since $c$ is negative, $|c|$ should be less than $(1+\epsilon)\sqrt{(1+\epsilon)^2-q(q+2)}$. 
Thus we have
\begin{equation}
c^2-\left[(1+\epsilon)\sqrt{(1+\epsilon)^2-q(q+2)}\right]^2
=2(1+\epsilon)^2q^2-2\epsilon^2(1+\epsilon)q-\epsilon(2\epsilon^2+3\epsilon+2)<0.
\nonumber
\end{equation}
In order that the above inequality folds, $q$ should satisfy
\begin{equation}
0<q<q_*, \nonumber
\end{equation}
where
\begin{equation}
q_*:=\frac{\epsilon^2+\sqrt{\epsilon(1+\epsilon)(2+\epsilon+\epsilon^2)}}{2(1+\epsilon)}. \nonumber
\end{equation}
Since we are interested in the case of $0<\epsilon\ll1$, we have 
$q_*=\sqrt{\epsilon/2}+{\cal O}(\epsilon^{3/2})<\sqrt{1+(1+\epsilon)^2}-1$. 
In order to know which is larger, $b$ or $b_-$, we should again note that 
$b$ is the larger root of $A_1(b;q)=0$, whereas $b_\pm$ are the roots of  
$G(b_\pm;q)=0$ with $b_+>b_-$. We should also note that 
there is only one root of the equation $A_1(\mu;q)=G(\mu;q)$: it is given by
\begin{equation}
\mu=b_{\rm m}:=\frac{(1+\epsilon)^2q(q+2)}{1+(1+\epsilon)^2}. \nonumber
\end{equation}
The function $A_1(\mu;q)$ takes a minimum value at $\mu=-1$, whereas 
the function $G(\mu;q)$ takes a minimum value at $\mu=(1+\epsilon)^2$. 
Thus, we have $-1<b_{\rm m},~b,~b_-<(1+\epsilon)^2$. 
From these facts, we can see that $b<b_{\rm m}<b_-$ for 
$A_1(b_{\rm m};q)=G(b_{\rm m};q)>0$, 
$b=b_{\rm m}=b_-$ for $A_1(b_{\rm m};q)=G(b_{\rm m};q)=0$,  
and $b>b_{\rm m}>b_-$ for $A_1(b_{\rm m};q)=G(b_{\rm m};q)<0$. 
We have
\begin{equation}
A_1(b_{\rm m};q)=G(b_{\rm m};q)=\frac{(1+\epsilon)^2q(q+2)}{[1+(1+\epsilon)^2]^2}
\left[(1+\epsilon)^2q(q+2)-\left\{(1+\epsilon)^4-1\right\}\right]. \nonumber
\end{equation}
From the above equation, we can see that  
$A_1(b_{\rm m};q)=G(b_{\rm m};q)>0$ for $q>q_2$, 
$A_1(b_{\rm m};q)=G(b_{\rm m};q)=0$ for $q=q_2$,  
and $A_1(b_{\rm m};q)=G(b_{\rm m};q)<0$ for $0<q<q_2$ (note that 
we focus on the case of $q>0$.), where  
\begin{equation}
q_2:=1+\frac{\sqrt{(1+\epsilon)^4+(1+\epsilon)^2-1}}{1+\epsilon}. \nonumber
\end{equation}
Hence, we have
\begin{eqnarray}
b&>&b_-~~~~~{\rm for}~~0<q<q_2, \cr
b&=&b_-~~~~~{\rm for}~~~~~q=q_2, \cr
b&<&b_-~~~~~{\rm for}~~~~~q>q_2, \nonumber
\end{eqnarray}
In the case of $0<\epsilon\ll1$, we have $q_2=2+{\cal O}(\epsilon)$ and hence 
we find that $q_*<q_2$. This implies that $b>b_-$ holds for $0<q<q_*$. 
Thus, Eq.~(\ref{ineq-6}) should be replaced by
\begin{eqnarray}
b<\mu<q(1+\epsilon)+\epsilon~~~~~~~{\rm for}~~~~0<q< q_*. \label{not-sharp}
\end{eqnarray}

Next, we consider the condition (\ref{V2-seigen}). 
In order that $r=r_{\rm min}$ is in the allowed domain for 
the motion of the shell-2, $H(\mu;q)$ should be 
non-negative. This condition is written in the form
\begin{equation}
|I(\mu;q)|>2\mu(1+\epsilon)\sqrt{(1+\epsilon)^2-1}, \label{con-0}
\end{equation}
where
\begin{equation}
I(\mu;q):=\mu^2+2(1+\epsilon)^2\mu-(1+\epsilon)^2q(q+2).
\end{equation}

\subsubsection{The case with negative charge or without charge}

In this case, the condition (\ref{ineq-5}) leads to $q(q+2)<0$. Then, since $\mu>0$, 
we can easily see $I(\mu;q)>0$. 
The condition (\ref{con-0}) is written in the form
\begin{equation}
\mu^2+2(1+\epsilon)\left[1+\epsilon-\sqrt{(1+\epsilon)^2-1}\right]-(1+\epsilon)^2q(q+2)>0.
\end{equation}
The above condition is always satisfied under the condition (\ref{ineq-5}), since 
$q(q+2)<0$ and $\mu>0$. 
As a result, the sphere $r=r_{\rm min}$ is in the allowed domain for the motion of the shell-2 
under the condition (\ref{ineq-5}).

\subsubsection{The case with positive charge}

We consider the case of $q>0$, and Eq.~(\ref{not-sharp}) should be satisfied. 
Here note again that $A_1(b;q)=0$, and hence we can easily see that 
\begin{equation}
H(b;q)=-4b^2\epsilon(2+\epsilon)<0.
\end{equation}
Hence, in the case of $\mu=b$, the shell-2 cannot arrive at $r=r_{\rm min}$. We have 
\begin{eqnarray}
&&I(\mu;q)\leq0~~~~~~{\rm for}~~~~d_-\leq \mu \leq d_+, \\
&&I(\mu;q)>0~~~~~~{\rm for}~~\mu<d_-,~~ d_+<\mu,
\end{eqnarray}
where $d_\pm$ are the roots of $I(d_\pm;q)=0$: 
\begin{equation}
d_\pm=-(1+\epsilon)^2\pm\sqrt{(1+\epsilon)^4+(1+\epsilon)^2q(q+2)}.
\end{equation}
Since $d_-<0<b$ and $f(b)+2b\epsilon(2+\epsilon)>0$, we have $d_-<0<d_+<b$. This means that  
$I(\mu;q)>0$ for $\mu>b$.
Thus, the condition (\ref{con-0}) leads to 
\begin{eqnarray}
J(\mu;q)&:=&I(\mu;q)-2\mu(1+\epsilon)\sqrt{(1+\epsilon)^2-1} \cr
&=&
A_1(\mu;q)-2\mu\left[1+\epsilon-\sqrt{(1+\epsilon)^2-1}\right]\sqrt{(1+\epsilon)^2-1}>0. \label{con-1}
\end{eqnarray}
Together with $\mu>0$, the above condition leads to $\mu>\mu_{\rm m}$, where
\begin{equation}
\mu_{\rm m}:=(1+\epsilon)\left\{1+\epsilon-\sqrt{(1+\epsilon)^2-1}\right\}
\left[
\sqrt{1+\frac{q(q+2)}{\left\{1+\epsilon-\sqrt{(1+\epsilon)^2-1}\right\}^2}}-1
\right]. \nonumber
\end{equation}
Since $A_1(q(1+\epsilon)+\epsilon;q)=\epsilon(2+\epsilon)$, we have
\begin{eqnarray}
J(q(1+\epsilon)+\epsilon;q)&=&
\epsilon\left[(1+2\epsilon)(2+\epsilon)-2(1+\epsilon)\sqrt{(1+\epsilon)^2-1}\right] \cr
&-&2q(1+\epsilon)\left[1+\epsilon-\sqrt{(1+\epsilon)^2-1}\right]\sqrt{(1+\epsilon)^2-1}. \nonumber
\end{eqnarray}
From the above equation and Eq.~(\ref{con-1}), we obtain
\begin{equation}
q<q_2:=\frac{\epsilon\left[(1+2\epsilon)(2+\epsilon)-2(1+\epsilon)\sqrt{(1+\epsilon)^2-1}\right]}
{2\left[1+\epsilon-\sqrt{(1+\epsilon)^2-1}\right]\sqrt{(1+\epsilon)^2-1}}
=\left(\frac{\epsilon}{2}\right)^{1/2}\left(1+\frac{3}{4}\epsilon+{\cal O}(\epsilon^{3/2})\right). \nonumber
\end{equation}
Since $q_*=(\epsilon/2)^{1/2}[1-\epsilon/4+{\cal O}(\epsilon^{3/2})]$, we have 
$q_2>q_*$. Hence, instead of Eq.~(\ref{not-sharp}), the following condition should hold; 
\begin{equation}
\mu_{\rm m}<\mu<(1+\epsilon)q+\epsilon~~~~~{\rm for}~~0<q<q_*. \nonumber
\end{equation}


\begin{thebibliography}{99}
\bibitem{mass-parameters}
As already mentioned, 
Eq.~\eq{j-con-1} implies $f_{A+1}<f_A$, and 
this condition is rewritten in the form
$
{\cal M}_{A+1}(r)-{\cal M}_A(r)>0,
$
where ${\cal M}_A$ is equal to the so called Misner-Sharp mass of this system,   
$
{\cal M}_A(r)= M_A-Q_A^2/2r. 
$
The Misner-Sharp mass ${\cal M}(r)$ has a meaning of the energy within a sphere 
of radius $r$\cite{MS-mass, Kodama, Hayward}. 
Hence ${\cal M}_{A+1}(r)-{\cal M}_A(r)$ 
is regarded as the energy of the shell-$A$ of radias $r$ and 
is necessarily positive as expected, 
even if $M_{A+1}-M_A$ is negative.
\bibitem{innerblackhole}
Here it should be noted that Eq.~\eq{j-con-1-2} is derived 
from not only Eq.~\eq{j-con-1-1} but also
\begin{equation}
\sqrt{\dot{r}^2+f_{A+1}(r)}=-\sqrt{\dot{r}^2+f_A(r)}+\frac{m_A}{r}. \label{j-con-2}
\end{equation}
The above equation can be satisfied only if the shell 
exits in a black hole domain, i.e., $f_{A+1}<0$,
since $\dot{t}_{A+1}$ can be negative there.
By the similar argument as in the case of Eq.~(\ref{j-con-1}), 
the above equation leads to the following inequalities:
\begin{eqnarray}
r&\leq& \rho_A~~~~{\rm for}~~{\cal E}_A>0, \label{seigen-BH}\\
Q_{A+1}^2-Q_A^2+m_A^2&\geq&0~~~~~~{\rm for}~~~{\cal E}_A=0,   \\
r&\geq& \rho_A~~~~{\rm for}~~{\cal E}_A<0. 
\end{eqnarray}
In the case of ${\cal E}_A>0$, $Q_{A+1}^2-Q_A^2+m_A^2$ should be positive so that 
Eq.~(\ref{seigen-BH}) is satisfied.
\end{thebibliography}
\end{document}